
\magnification=1200
\baselineskip=12pt
\overfullrule=0pt
\font\titlea=cmb10 scaled\magstep1

\font\it=cmti10

\rightline{MRI-PHY 7/94, SBNC 06/94}
\baselineskip=18pt
\vskip .5cm
\centerline{\titlea N = 2 SUPER $W_{\infty}$ ALGEBRA AND ITS NONLINEAR
REALIZATION }
\medskip
\centerline{\titlea  THROUGH SUPER KP FORMULATION }
\vskip 2.5cm
\baselineskip=14pt
\centerline{\bf Sasanka Ghosh$^1$ and Samir K. Paul$^2$}
\bigskip
\centerline{1. {\it Mehta Research Institute of mathematics and
Mathematical Physics}}
\centerline{{\it 10, Kasturba Gandhi Marg, Allahabad 211 002, India}}
\bigskip
\centerline{2. {\it S N Bose National Centre for Basic Sciences}}
\centerline{{\it DB 17, Sector I, Salt Lake, Calcutta 700 064, India}}
\bigskip
\vskip 1.5cm
\baselineskip=18pt
\centerline{\titlea Abstract}
\bigskip
A nonlinear realization of super $W_{\infty}$ algebra is shown to exist
through a consistent superLax formulation of super KP hierarchy. The
reduction of the superLax operator gives rise to the Lax operators for
$N=2$ generalized super KdV hierarchies, proposed by Inami and Kanno.
The Lax equations are shown to be Hamiltonian and the associated Poisson
bracket algebra among the superfields, consequently, gives rise to a
realization of nonlinear super $W_{\infty}$ algebra.

\vfill
\eject

It has been known since the inception of Zamolodchikov, Fateev and
Lykyanov's [1] nonlinear realization of extended conformal symmetry that
$W_n$ algebra incorporates in its calssical limit the Hamiltonian
structure of nonlinear integrable systems {\it viz.}, generalized KdV
hierarchy [2]. Subsequently, there were important attempts to classify
conformal field theory via the Hamiltonian structure of generalized KdV
hierarchies [3]. The role of the generalized KdV hierarchies became
apparent later when it was seen that equations of motion and symmetries in
quantum 2D gravity and noncritical string theory at least with $c<1$ can
be formulated in terms of integrable nonlinear equations of KdV type.
(There is an exhustive literature in this direction; for example see [4]).
Since all the KdV hierarchies are contained into the larger integrable
system {\it viz.}, KP hierarchy [5] it has been conjectured [6] that it
may provide a universal framework exhibiting the underlying structure of
2D quantum gravity. In this connection and also from the point of view Lie
algebra, the algebraic structure of the large $n$ limit of $W_n$ algebra
namely, $W_{1+\infty}$ and $W_{\infty }$ has been probed within
the framework of KP hierarchy [7]. These are universal W algebras containg
all the conformal spins. All these algebras have so far been linear Lie
algebras. However, owing to the ambiguity inherent in the large $n$
limits, a nonlinear realization of universal W algebra, $\hat W_{\infty}$
has been constructed and identified with the second Hamiltonian structure
of KP hierarchy [8].

Later on, Manin and Radul [9] provided a consistent supersymmetric
extention of the KP hierarchy and subsequently the physically relevant
even parity superLax formulation of super KP hierarchy containing
linear universal W algebra (super $W_{1+\infty }$) has been developed
[10]. While linear $N=2$ universal super W algebra [10] was shown to
exist, Inami and Kanno [11, 12] have shown that $N=2$ super KdV
hierarchies
can arise in association with affine Lie algebras. This is an important
step towards $N=2$ super analogue of Drinfeld Sokolov formulation [2].
Clearly, it hints that, as a generalization of Inami and Kanno's work
[11, 12], there
must be a consistent $N=2$ superLax formulation of super KP hierarchy
which should be Hamiltonian with respect to super Gelfand Dikii bracket of
second kind and should reduce to the Lax operators, considered in [12]
under siutable reduction.
This observation, in fact, enables us to show the existence of a nonlinear
realization of super $W_{\infty }$ algebra through super KP formulation.

In this letter we formulate an even parity superLax operator which, under
suitable reductions, becomes isomorphic to the Lax operators of $N=2$
super KdV hierarchies [12]. Subsequently, we show that a consistent
superLax formulation
corresponding to this superLax operator leads to $N=2$ super KP
hierarchy. We also show the Lax euqations are Hamiltonian. For this
purpose, we obtain the Poisson bracket algebra among the coefficient
fields
corresponding to the superLax operator following Gelfand Dikii method.
We observe that the full Poisson bracket algebra has $N=2$ superconformal
algebra as a subalgebra and the bosonic and fermionic componets of the
superfields carry correct conformal weights. As a consequence, we claim
that the full Poisson bracket algebra among the superfields gives rise to a
realization of $N=2$ nonlinear super $W_{\infty }$ algebra.

We begin with the even parity superLax operator of the form
$$L = D^2 + \sum^{\infty }_{i=0} u_{i-1}(X) D^{-i} \eqno (1)$$
where, $D$ is the superderivative with $D^2 = {d\over {dx}}$ and $u_{i-1}(X)$
are superfields in $X = (x,\theta )$ space, $\theta $ being Grassman odd
coordinate. The grading of $u_{i-1}(X)$ is $\vert u_{i-1}\vert = i$ so that
$u_{2i-1}$ are bosonic superfields whereas $u_{2i}$ are fermionic ones.

The interesting consequence of the choice of the superLax operator (1) is
that under suitable choice of reduction it reduces to the superLax
operator for generalized $N=2$ KdV hierarchies, proposed by Inami and
Kanno. Let us first define the n{\it th.} reduction of $L$ as
$${\tilde L}_n = L^n_{> 0} \eqno (2)$$
\noindent where, `$> 0$' implies the +ve part of $L^n$ without $D^0$
term. Using definition (2) for the superLax operator (1), $\tilde L_n$ can
be expressed in the following general form
$${\tilde L}_n = L^n_{> 0} = D^{2n} + \sum^{2n-2}_{i=1} {\cal U}^{(n)}_i
D^i  \eqno (3)$$
\noindent In (3) we have used super Liebnitz rule [9] for multiplication
of the operators $L$ and the superfields ${\cal U}^{(n)}_i$ are functions
of $u_{i-1}(X)$ and their superderivatives. $\tilde L_n$ in (3) is
precisely the superLax operator considered in [12]. For example, for $n=2$
$$\tilde L_2 = D^4 + 2 u_{-1} D^2 + 2 u_0 D \eqno (4)$$
\noindent which is precisely $N=2$ super KdV Lax operator [11]. Similarly,
for $n=3$
$${\tilde L}_3 = D^6 + 3 u_{-1} D^4 + 3 u_0 D^3 + 3 (u_1 + u_{-1}^{[2]} +
u_{-1}^2) D^2 + 3 (u_2 + u_0^{[2]} + 2 u_{-1} u_0) D \eqno (5)$$
\noindent is isomorphic to the Lax operator corresponding to $N=2$ super
Boussinesq hierarchy [11]. Notice that we have chosen the reduction
prescription as in definition (2), since $D^0$ term is absent in the
superLax operators (3, 4, 5), considered in [11, 12]. This observation is,
in fact,
a compelling evidence to propose the superLax operator in (1) as a right
candidate for describing $N=2$ super KP hierarchy.

It is clear from (4, 5) that the presence of the superfield $u_{-1}$ in (1) is
essential to obtain superLax operators for $N=2$ generalized
super KdV
hierarchies by reduction. Hence $u_{-1}$ must have nontrivial dynamics.
In order to have nontrivial flow of $u_{-1}$ for each time $t_n$  we now
define the Lax equation as
$${{dL}\over {dt_n}} = \left[ L^n_{> 0} , L \right] \eqno (6)$$
We will show later that the presence of $u_{-1}$ renders the Gelfand Dikii
Poisson bracket of second kind local whereas the absence of $u_{-1}$ field
makes the same nonlocal [13].

We give below first three evolution equations which follow from (1) and
(6).
$$\eqalignno{ {du_{i-1}\over {dt_1}} = & u_{i-1}^{[2]} &(7a)\cr
{du_{i-1}\over {dt_2}} = & 2 u_{i+1}^{[2]} + u_{i-1}^{[4]} + 2 u_{-1}
u_{i-1}^{[2]} + 2 u_0 u_{i-1}^{[1]} - 2 \left[{i+1\atop 1}\right] u_i
u_{-1}^{[1]} - 2 (1 + (-1)^i) u_0 u_i \cr & + 2 \sum ^{i-1}_{m=0}
\left[{i\atop {m+1}}\right] (1-)^{i+[-{m\over 2}]} u_{i-m-1} u_0^{[m+1]}
\cr & +
2 \sum ^{i-1}_{m=0} \left[{i+1\atop {m+2}}\right] (1-)^{[{m\over 2}]}
u_{i-m-1} u_{-1}^{[m+2]} &(7b)\cr
{du_{i-1}\over {dt_3}} = & 3 u_{i+3}^{[2]} + 3 u_{i+1}^{[4]} +
u_{i-1}^{[6]} + 6 u_{-1} u_{i+1}^{[2]} + 3 u_{-1} u_{i-1}^{[4]} \cr &
- 3 \left[{i+3\atop 1}\right] u_{i+2} u_{-1}^{[1]} + 3 \left[{i+3\atop
2}\right] u_{i+1}u_{-1}^{[2]} + 3 \left[{i+3\atop 3}\right] u_i
u_{-1}^{[3]} \cr & - 3 (1 + (-1)^i) u_0 u_{i+2} + 3 u_0 u_{i+1}^{[1]} - 3
(-1)^i u_0u_i^{[2]} + 3 u_0 u_{i-1}^{[3]} \cr & + 3 \left[{i+2\atop
1}\right] (-1)^i u_{i+1}u_0^{[1]} - 3 \left[{i+2\atop 2}\right] (-1)^i u_i
u_0^{[2]} \cr & + 3 (u_1 + 2 u_{-1}^{[2]} + u_{-1}^2) u_{i-1}^{[2]} - 3
\left[{i+1\atop 1}\right] u_i (u_1 + u_{-1}^{[2]} + u_{-1}^2)^{[1]} \cr &
+ 3 (u_2 + 2 u_{-1} u_0 + u_0^{[2]}) u_{i-1}^{[1]} - 3 (1 + (-1)^i)
(u_2 + 2 u_{-1} u_0 + u_0^{[2]}) u_i \cr &
- 3 \sum ^{i-1}_{m=0} \left[{i+3\atop {m+4}}\right] (-1)^{[{m\over 2}]}
u_{i-m-1} u_{-1}^{[m+4]} \cr &
- 3 \sum ^{i-1}_{m=0} \left[{i+2\atop {m+3}}\right] (-1)^{i+[-{m\over 2}]}
u_{i-m-1} u_0^{[m+3]} \cr &
+ 3 \sum ^{i-1}_{m=0} \left[{i+1\atop {m+2}}\right] (-1)^{[{m\over 2}]}
u_{i-m-1} (u_1 + u_{-1}^{[2]} + u_{-1}^2)^{[m+2]} \cr &
+ 3 \sum ^{i-1}_{m=0} \left[{i\atop {m+1}}\right] (-1)^{i+[-{m\over 2}]}
u_{i-m-1} (u_2 + 2 u_{-1} u_0 + u_0^{[2]})^{[m+1]} &(7c) \cr}$$
With the identification of $t_1 = x$ and $t_2 = y$, (7a) resembles the
consistency condition, whereas (7b) becomes constraint equation. The time
variables, therefore, may be identified as $t_3, t_4,....$ {\it etc.}

Let us now look for the $t_3$ time evolution (first time evolution)
equations for the superfields $u_{-1}$ and $u_0$. We may eliminate all
other fields from the equations of motion of $u_{-1}$ and $u_0$ in (7c) by
using the constraint (7b). As a consequence, however, the equations of
motion for $u_{-1}$ and $u_0$ become nonlocal and have the form
$$\eqalignno{{du_{-1}\over {dt_3}} &= {1\over 4} u_{-1}^{[6]} - {1\over 2}
(u^3_{-1})^{[2]} + {3\over 2} (u_0 u_{-1}^{[1]})^{[2]} + {3\over 4}
{d^2u_{-1}^{[-2]}\over {dy^2}} \cr
&+ {3\over 2} u_{-1}^{[2]} {du_{-1}^{[-2]}\over {dy}} - {3\over 2}
u_{-1}^{[1]} {du_0^{[-2]}\over {dy}} - 3 u_0 {du_{-1}^{[-1]}\over {dy}} +
3 u_0 {du_0^{[-2]}\over {dy}} &(8a) \cr
{du_0\over {dt_3}} &= {1\over 4} u_0^{[6]} + {3\over 2}
(u_0 u_0^{[1]})^{[2]} - {3\over 2} (u_0 u_{-1}^{[2]})^{[2]} - {3\over 2}
(u_0 u_{-1}^2)^{[2]} \cr &+ {3\over 4}
{d^2u_0^{[-2]}\over {dy^2}} + {3\over 2} (u_0 {du_0^{[-2]}\over
{dy}})^{[1]} + {3\over 2} u_0^{[2]} {du_{-1}^{[-2]}\over {dy}} + {3\over 2}
u_0 {du_{-1}\over {dy}} &(8b) \cr}$$
The higher time evolution of this hierarchy also have similar nonlocal
terms involving $y$ derivate only.

If we further demand that the superfields are independent of $y$
coordinate the evolution equations of $u_{-1}$ and $u_0$ become local and
reduce to the equations,
$${du_{-1}\over {dt_3}} = - \left[ u_{-1}^{[6]} + 3 (u_{-1}^{[1]}
u_0)^{[2]} - {1\over 2} (u_{-1}^3)^{[2]} \right] \eqno (9a)$$
and
$${du_0\over {dt_3}} = - \left[ u_0^{[6]} - 3 (u_0 u_0^{[1]})^{[2]} -
{3\over 2} (u_0 u_{-1}^2)^{[2]} + 3 (u_0 u_{-1}^{[2]})^{[2]} \right] \eqno
(9b)$$
\noindent after rescaling of $u_{-1} = - {1\over 2} u_{-1}, u_0 = -
{1\over 2} u_0$ and $t_3 = - {1\over 4} t_3$. (9) are the evolution
equations for $N=2$ super KdV system [11].

Moreover, we show that (7) contain KP equation in the bosonic limit. For
this purpose, let us first write down the superfields in the component
forms as
$$\eqalignno{u_{2i-1}(X) =& u_{2i-1}^b(x) + \theta u_{2i-1}^f(x)
&(10a)\cr u_{2i}(X) =& u_{2i}^f(x) + \theta u_{2i}^b(x) &(10b)\cr}$$
\noindent where, $b$ and $f$ denote fermion and boson respectively. In
particular, it follows from (7) that the equation of motion of $u_0^b(x)$
have the following form
$${3\over 4} {{d^2 u_0^b(x)}\over {dy^2}} = {d\over {dx}} \left(
{{du_0^b(x)}\over {dt_3}} - {1\over 4} {{d^3u_0^b(x)}\over {dx^3}} - 3
u_0^b(x) {{du_0^b(x)}\over {dx}} \right) \eqno (11)$$
\noindent after setting the fermionic components of the superfields and
$u_{-1}^b(x)$ field to zero. (11) is, indeed, the KP equation [14]. Thus
the set
of equations (7) is nothing but the super KP equation. In addition, we
have shown that (7) reduce to $N=2$ super KdV equations (9). Hence, this
suggests that (1) and (6) describe the dynamics of $N=2$ super KP
hierarchy. Now, it remains to show that the Poisson bracket algebra among
the superfields corresponding to (1) and (6) has $N=2$ superconformal
algebra as a subalgebra.

To show that the Lax equations (6, 7) are Hamiltonian, we first calculate
the Poisson bracket algebra among the coefficient fields $u_{i-1}(X)$
following the method of Gelfand and Dikii. The super Gelfand Dikii bracket
is defined as
$$\left\{ F_P(L) , F_Q(L) \right\} = - Tr \left[ \left\{ L(PL)_- -
(LP)_-L \right\} Q \right] \eqno (12)$$
\noindent where, $P, Q$ are auxiliary fields (analogous to the super
Volterra operators in the case of super KdV system). We choose $P, Q$ in
the form
$$P = \sum_{j=-2}^{\infty }D^j p_j \qquad ; \qquad Q = \sum_{j=-2}^{\infty
}D^j q_j \eqno (13)$$
\noindent with the grading $\vert p_j \vert = \vert q_j \vert = j$ so that
the linear functional $F_P(L)$ (and similarly $F_Q(L)$) becomes
$$F_P(L) = Tr (LP) = \sum_{i=0}^{\infty } \int dX (-1)^{i+1} u_{i-1}(X)
p_{i-1}(X) \eqno (14)$$
Consequently the L.H.S. of (12) becomes
$$\left\{ F_P(L) , F_Q(L) \right\} = \sum_{i,j=0}^{\infty } \int dX
\int dY (-1)^{i+j} p_{i-1}(X) \left\{ u_{i-1}(X) , u_{j-1}(Y) \right\}
q_{j-1}(Y) \eqno (15)$$
Notice that (15) does not involve terms like $p_{-2}$ and $q_{-2}$ since
the superfields in (1) starts from $u_{-1}(X)$. To ensure this consistency
we have to show that R.H.S. of (12) sets the coefficients of $p_{-2}$ and
$q_{-2}$ identically zero. This needs the coefficient of $D$ term in the
$V_P(L)$ to be zero, where $V_P(L)$ is defined by
$$V_P(L) = L(PL)_- - (LP)_-L. $$
The above condition thus leads to the constraint
$$p_{-1}(X) = - \sum_{r=0}^{\infty } \sum_{m=0}^{r-1} \left[ {r -1\atop {m
+1}} \right] (-1)^{m(r+1)} \left( p_{r-1}(X) u_{r-m-2}(X) \right)^{[m-1]}
\eqno (16)$$
In particular, vanishing of the coefficient of $D$ term ensures that
R.H.S. of (12) is independent of $q_{-2}$. Now by using the constraint
(16) we can make the coefficient of $p_{-2}$ zero. We remark that the
origin of the constraint (16) is due to the absence of $D$ term in the
superLax operator (1) itself. Finally, we obtain the Poisson bracket
algerba among the superfields by using the relations (12)-(16). Thus we
have
$$\eqalign{ &\left\{ u_{j-1}(X) , u_{k-1}(Y) \right\} = \cr
&\left[ - \sum_{m=0}^{j+1} \left[{j + 1\atop m}\right] (-1)^{j(k+m+1) +
[{m\over 2}]} u_{j+k-m} D^m \right. \cr
&+ \sum_{m=0}^{k+1} \left[{k + 1\atop m}\right] (-1)^{jm+(k+1)(m+1)} D^m
u_{j+k-m} \cr
&+ \sum_{m=0}^{j-1} \sum_{l=0}^{k-1} \left( \left[{j\atop {m + 1}}\right]
\left[{k\atop {l + 1}}\right] - \left[{j - 1\atop m}\right]
\left[{k - 1\atop l}\right] \right) (-1)^{j(m+1)+k+l+[{m\over 2}]} \cr
&~~~~~~~~~~~~~~~~~~~~~~~~~~~~~~~~~~~~~~~~~u_{j-m-2} D^{m+l+1} u_{k-l-2} \cr
&+ \sum_{n=0}^{k-1} \sum_{l=0}^{k-n-1} \left[{k - n - 1\atop l}\right]
(-1)^{j(n+l)+(l+1)(n+k+1)} u_{n-1} D^l u_{j+k-n-l-2} \cr
&- \sum_{m=0}^{j+k-n-l-1} \sum_{n=0}^{k-1} \sum_{l=0}^{k-n-1}
\left[{j - 1\atop m}\right] \left[{n +l - 1\atop l}\right]
(-1)^{j(m+n+l+k+1)+n(l+1)+[{m\over 2}]} \cr
&~~~~~~~~~~~~~~~~~~~~~~~~~~~~~~~~~~\left. u_{j+k-m-n-l-2} D^{m+l} u_{n-1}
\right] \Delta (X -Y)} \eqno (17)$$
Next we define the Hamiltonians, $H_n$ as
$$H_n = {1\over n}\int dX sRes (L^n)(X) \eqno (18)$$
\noindent for $n=1,2,3,...$ etc. Here $`sRes'$ means superresidue,{\it
i.e.} the coefficient of $D^{-1}$. To check that first few equations, (7)
satisfy Hamilton's equation
$${du_{i-1}(X)\over {dt_n}} = \left\{ u_{i-1}(X) , H_n \right\} \eqno (19)$$
we give explicit form of first three Hamiltonians
$$\eqalign{H_1 =& \int dX u_0 \cr H_2 =& \int dX (u_2 + u_{-1} u_0) \cr
H_3 =& \int dX (u_4 + 2 u_2 u_{-1} + 2 u_1 u_0 + u_0 u_0^{[1]} + u_{-1}
u_0^{[2]} + u_0 u_{-1}^2) \cr} \eqno (20)$$
For example, for $n=1$, (19) becomes
$${du_{i-1}(X)\over {dt_1}} = (-1)^i \int dY \left\{ u_{i-1}(X) , u_0(Y)
\right\} \eqno (21)$$
Now by using (17) we have from (21)
$${du_{i-1}(X)\over {dt_1}} = u_{i-1}^{[2]}(X)$$
\noindent which is nothing but (7a). Similarly for $n=2,3$ we have
verified by using (17), (19) and (20) that Hamilton's equations exactly
match with (7b,c).

A few remarks about the Poisson bracket algebra (17) are in order.

\noindent (i) \quad The first three terms in algebra are manifestly
antisymmetric,
but the last two terms are not apparently antisymmetric. This causes
obstruction in proving Jacobi identity in a covariant fashion by using the
manifestly antisymmetry property  and cyclicity in the indices. We have,
however, checked for $k=0,1,2,3$ and arbitrary  $j$ and also for $j=0,1,2,3$
and arbitrary $k$ that Poisson brackets are, indeed, antisymmetric.
For convenience, we display the explicit expressions of these Poisson brackets
below. From (17) the Poisson brackets between $k=0,1,2,3$ and arbitrary
$j$ are given by
$$\eqalignno{ \left\{ u_{j-1}(X) , u_{-1}(Y) \right\} &= \left( - u_j +
(-1)^j D u_{j-1} \right. \cr
& \left. - \sum_{m=0}^{j+1} \left[{j + 1\atop m}\right]
(-1)^{j(m+1)+[{m\over 2}]} u_{j-m} D^m \right) \Delta (X - Y) &(22a) \cr
\left\{ u_{j-1}(X) , u_0(Y) \right\} &= \left(
D^2 u_{j-1} \right. \cr & \left. - \sum_{m=0}^j
\left[{j + 1\atop {m + 1}}\right]
(-1)^{j(m+1)+[-{m\over 2}]} u_{j-m} D^{m+1} \right) \Delta (X - Y) &(22b)
 \cr}$$
$$\eqalign{& \left\{ u_{j-1}(X) , u_1(Y) \right\} = \left( - u_{j+2} +
(-1)^j D u_{j+1} - D^2 u_j +(-1)^j D^3 u_{j-1} + (-1)^j u_0 u_{j-1}
\right.\cr & - u_{-1} u_j - u_{j-1} D u_{-1} + (-1)^j u_{-1} D u_{j-1}
- \sum_{m=0}^{j+1} \left[{j + 1\atop m}\right]
(-1)^{j(m+1)+[{m\over 2}]} u_{j+2-m} D^m
 \cr & \left.
- \sum_{m=0}^j \left[{j\atop m}\right] (-1)^{j(m+1)+[-{m\over 2}]}
\left( u_{j-m} D^m u_{-1}
- (-1)^j u_{j-m-1} D^m u_0 \right) \right) \Delta (X - Y)} \eqno(22c)$$
$$\eqalign{& \left\{ u_{j-1}(X) , u_2(Y) \right\} = \left(
2 D^2 u_{j+1} + D^4 u_{j-1} - (-1)^j u_0 u_j +u_0 D u_{j-1} - u_{j-1} D^2
u_{-1} \right.\cr & + u_{-1} D^2 u_{j-1}
 - \sum_{m=0}^j \left[{j + 1\atop {m + 1}}\right]
(-1)^{[-{m\over 2}]} \left( u_{j+2-m} D^{m+1} + u_{j-m} D^{m+1} u_{-1}
 \right) \cr & \left.
+ \sum_{m=0}^j \left[{j\atop m}\right] (-1)^{j+[{m\over 2}]}
u_{j-m} D^m u_0 - \sum_{m=0}^{j-1} \left[{j\atop {m + 1}}\right]
(-1)^{[-{m\over 2}]} u_{j-m-2} D^{m+2} u_0 \right) \Delta (X - Y)}
\eqno(22d)$$
It also follows from (17) that the Poisson brackets between $j=0,1,2,3$ and
arbitrary $k$ have the form
$$\eqalignno{ \left\{ u_{-1}(X) , u_{k-1}(Y) \right\} &= \left( - u_k
- u_{k-1} D \right. \cr
& \left. + \sum_{m=0}^{k+1} \left[{k + 1\atop m}\right]
(-1)^{(k+1)(m+1)} D^m u_{k-m} \right) \Delta (X - Y) &(23a) \cr
\left\{ u_0(X) , u_{k-1}(Y) \right\} &= \left(
- (-1)^k u_{k-1} D^2 \right. \cr & \left.
- \sum_{m=0}^k \left[{k + 1\atop {m + 1}}\right]
(-1)^{km} D^{m+1} u_{k-m} \right) \Delta (X - Y) &(23b) \cr}$$
$$\eqalign{& \left\{ u_1(X) , u_{k-1}(Y) \right\} = \left( - u_{k+2} -
u_{k+1} D + u_k D^2 + u_{k-1} D^3 + u_{k-1} u_0 \right.\cr &
- u_k u_{-1} - u_{k-1} D u_{-1} + (-1)^k u_{-1} D u_{k-1}
+ \sum_{m=0}^{k+1} \left[{k + 1\atop m}\right]
(-1)^{(k+1)(m+1)} D^m u_{k+2-m} \cr & \left.
- \sum_{m=0}^k \left[{k\atop m}\right] (-1)^k \left( u_{-1} D^m u_{k-m}
- (-1)^m u_0 D^m u_{k-m-1} \right) \right) \Delta (X - Y)} \eqno(23c)$$
$$\eqalign{& \left\{ u_2(X) , u_{k-1}(Y) \right\} = - (-1)^k \left(
2 u_{k+1} D^2 - u_{k-1} D^4 - u_k u_0 - u_{k-1} D u_0
+ u_{k-1} D^2 u_{-1} \right.\cr & - u_{-1} D^2 u_{k-1}
+ \sum_{m=0}^k \left[{k + 1\atop {m + 1}}\right] \left(
D^{m+1} u_{k+2-m} + u_{-1} D^{m+1} u_{k-m} \right) \cr & \left.
+ \sum_{m=0}^k \left[{k\atop m}\right] u_0 D^m u_{k-m}
- \sum_{m=0}^{k-1} \left[{k\atop {m + 1}}\right](-1)^m u_0 D^{m+2} u_{k-m-2}
\right) \Delta (X - Y)} \eqno(23d)$$
It is now easy to see that (22a) is antisymmetric to (23a) and so on. We have
also checked that Jacobi identities are satisfied  for the above cases.

\noindent (ii) \quad The algebra is local. This enables us to associate it
with a superconformal algebra. To make contact the algebra (17) with extended
$N=2$ superconformal algebra, {\it viz.} super $W_{\infty }$ algebra, we
first consider the Poisson bracket algebra between the superfields
$u_{-1}(X)$ and $u_0(X)$. From (17) we have
$$\eqalign{ \left\{ u_{-1}(X) , u_{-1}(Y) \right\} & = \left( - 2 u_0(X) -
u_{-1}(X) D_X + D_X u_{-1}(X) \right) \Delta (X - Y) \cr
\left\{ u_0(X) , u_{-1}(Y) \right\} & = \left( - D_X u_0(X) - u_{-1}(X)
D_X^2 \right) \Delta (X - Y) \cr
\left\{ u_{-1}(X) , u_0(Y) \right\} & = \left( - u_0(X) D_X + D_X^2
u_{-1}(X) \right) \Delta (X - Y) \cr
\left\{ u_0(X) , u_0(Y) \right\} & = \left( u_0(X) D_X^2 + D_X^2
u_0(X) \right) \Delta (X - Y)} \eqno (24)$$
\noindent which is closed and hence is a subalgebra of the full algebra
(17). To show that (24) has $N=2$ superconformal structure it is useful to
write (24) in terms of component fields by using (10). If we further
redifine the fields as
$$T = u_0^b - {1\over 2} (u_{-1}^b)^{\prime}, \qquad U = u_{-1}^b, \qquad
G^+ = u_0^f, \qquad G^- = u_0^f - u_{-1}^f \eqno (25)$$
\noindent the algebra (24) in terms of $T, U, G^+$ and $G^-$ becomes
$$\eqalign{ \left\{ T(x) , T(y) \right\} &= \left( 2 T(y) \partial _y +
T^{\prime }(y) \right) \delta (x - y) \cr
\left\{ T(x) , U(y) \right\} &= \left( U(y) \partial _y + U^{\prime }(y)
\right) \delta (x - y) \cr
\left\{ T(x) , G^{\pm }(y) \right\} &= \left( {3\over 2} G^{\pm }(y)
\partial _y + (G^{\pm })^{\prime }(y) \right) \delta (x - y) \cr
\left\{ G^{\pm }(x) , U(y) \right\} &= \pm G^{\pm }(y) \delta (x - y) \cr
\left\{ G^+(x) , G^-(y) \right\} &= \left( T(y) - U(y)
\partial _y - {1\over 2} U^{\prime }(y) \right) \delta (x - y) \cr
\left\{ G^{\pm }(x) , G^{\pm }(y) \right\} &=
\left\{ U(x) , U(y) \right\} = 0} \eqno (26)$$
(26) is nothing but the classical analogue of $N=2$ superconformal algebra.

\noindent (iii) \quad Let us now calculate the Poisson bracket of the componet
fields $u_{2i-1}^b, u_{2i}^b, u_{2i-1}^f$ and $u_{2i}^f$ with the energy
momentum tensor T, defined in (25). It follows from (10) and (17) that
$$\eqalignno{ \left\{ T(x) , u_{2i-1}^b(y) \right\} &= \left( (i + 1)
u_{2i-1}^b(y) \partial _y + (u_{2i-1}^b(y))^{\prime } \right. \cr &
- \sum_{m=0}^{i-2}
(-1)^m \left( {i\atop {m + 2}} \right) u_{2i-2m-3}^b(y) \partial _y^{m+2}
\cr &\left. - {1\over 2} \sum_{m=0}^{i-1} (-1)^m \left( {i\atop {m
+ 1}} \right) u_{2i-2m-3}^b(y) \partial _y^{m+2} \right) \delta (x - y)
&(27a) \cr
\left\{ T(x) , u_{2i}^b(y) \right\} &= \left( (i + 2)
u_{2i}^b(y) \partial _y + (u_{2i}^b(y))^{\prime } \right. \cr &
- \sum_{m=0}^{i-2}
(-1)^m \left( {i + 1\atop {m + 2}} \right) u_{2i-2m-2}^b(y) \partial _y^{m+2}
\cr &\left. + {1\over 2} \sum_{m=0}^{i-1} (-1)^m \left( {i + 1\atop {m
+ 1}} \right) u_{2i-2m-1}^b(y) \partial _y^{m+2} \right) \delta (x - y)
&(27b) \cr
\left\{ T(x) , u_{2i-1}^f(y) \right\} &= \left( (i + {3\over 2})
u_{2i-1}^f(y) \partial _y + (u_{2i-1}^f(y))^{\prime } \right. \cr &
- \sum_{m=0}^{i-2}
(-1)^m \left( {i\atop {m + 2}} \right) u_{2i-2m-3}^f(y) \partial _y^{m+2}
\cr &\left. + {1\over 2} \sum_{m=0}^{i-1} (-1)^m \left( {i\atop {m
+ 1}} \right) u_{2i-2m-3}^f(y) \partial _y^{m+2} \right) \delta (x - y)
&(27c) \cr
\left\{ T(x) , u_{2i}^f(y) \right\} &= \left( (i + {3\over 2})
u_{2i}^f(y) \partial _y + (u_{2i}^f(y))^{\prime } \right. \cr & \left.
- \sum_{m=0}^{i-2}
(-1)^m \left( {i + 1\atop {m + 2}} \right) u_{2i-2m-2}^f(y) \partial _y^{m+2}
\right) \delta (x - y) &(27d) \cr}$$
It is now evident from (27) that $u_{2i-1}^b, u_{2i}^b$ and $u_{2i-1}^f,
u_{2i}^f$ are respectively bosonic and fermionic conformal fields with
respect to the energy momentum tensor T, introduced in (25).

Hence we cliam that the Poisson bracket algebra (17) is a nonlinear
realization of super $W_{\infty }$ algebra. Notice that similar situation
has been observed in the case of Gelfand Dikii bracket of second kind for
the bosonic KP hierarchy [8].

To conclude, the superLax operator (1), indeed, corresponds to $N=2$ super KP
hierarchy. The Poisson bracket algebra, we have obtained following Gelfand
Dikii method, gives rise to a nonlinear realization of super $W_{\infty }$
algebra. In alalogy with bosonic KP hierarchy [8] we believe this algebra
may be realized as universal super $W_{\infty }$ algebra.

{\it Authors would like to thank S. Panda for discussion in the early
stage of the work. One of us (SG) is thankful to S. Roy for discussion and
also to E. Ivanov and S. Krivonos for bringing the reference [11] to his
attention.}
\vfill\break
\noindent {\titlea References :}
\bigskip

\noindent 1. A.B. Zamolodchikov and V.A. Fateev, Nucl Phys. B280 [FS18]
(1987) 644; V.A. Fateev and S.L. Lykyanov, Int. J Mod. Phys. A3 (1988) 507.

\noindent 2. V.G. Drinfeld and V.V. Sokolov, Lie Algebra and Equations of
Korteweg-De Vries Type; 1985 Plenum Publishing Corporation p. 1975-2036.

\noindent 3. A.A. Belavin, Adv. Stud. Pure Math. 19 (1989) 117; I. Bakas,
Phys. Lett. B213 (1988) 313; I. Bakas, Phys. Lett. B219 (1989) 283;

\noindent 4. P.Di Francesco, P. Ginsparg and J. Zinn-Justin (2D Gravity
and Random Matrices, Saclay preprint LA-UR-93-1722, SPHT/93-061, (to
appear in Phys. Rep. C) and the references therein.

\noindent 5. M. Sato, RIMS Kokyuroku 439 (1981) 30; E. Date, M. Jimbo, M.
Kashiwara and T. Miwa, in Proc. RIMS Symp. on Nonlinear Integrable
Systems, eds. M. Jimbo and T. Miwa (World Scientific, 1983); G. Segal and
G. Wilson, Publ. IHES 61 (1985) 1.

\noindent 6. M. Fukuma, H. Kawai and R. Nakayama, Int. J. Mod. Phys. A6
(1991) 1385; M.A. Awada and S.J. Sin, Int. J. Mod. Phys. A7 (1992) 4791.

\noindent 7. C. Pope, L. Romans and
X. Shen, Nucl. Phys. B339 (1990) 191; Phys. Lett. B242 (1990) 401; I.
Bakas and E. Kiritsis, Nucl. Phys. B343 (1990) 185; K. Yamagishi, Phys.
Lett B259 (1991) 436; F. Yu and Y.S. Wu, Phys. Lett. B263 (1991) 220.

\noindent 8. A. Das, W-J Huang and S. Panda, Phys. Lett. B271 (1991) 109;
F. Yu and Y-S Wu, Nucl. Phys. B373, (1992) 713; I. Bakas and E. Kiritsis,
Int. J. Mod. Phys. A7, Suppl. 1A (1992) 55.

\noindent 9. Yu.I. Manin and A.O. Radul, Com. Math. Phys. 98 (1985) 65.

\noindent 10. E. Bergshoeff, C. Pope, L. Romans, E. Sezgin and X. Shen,
Phys. Lett. B245 (1990) 447; F. Yu, Nucl.Phys. B375 (1992) 173.

\noindent 11. T. Inami and H. Kanno, Nucl. Phys. B359 (1990) 185.

\noindent 12. T. Inami and H. Kanno, Int. J. Mod. Phys. A7, Suppl. 1A
(1992) 419 and the references therein.

\noindent 13. J. Barcelos-Neto, S. Ghosh and S. Roy, preprint IC/93/179.

\noindent 14. B.B. Kadomtsev and V.I. Petviashvili, Sov. Phys. Dokl. 15
(1971) 359.

\vfill\end